\documentclass[aps,prx,superscriptaddress,twocolumn,amsmath]{revtex4-1}
\usepackage{graphicx, color}
\usepackage{amsfonts,amssymb,amsmath}
\usepackage{verbatim}
\usepackage{bm}

\renewcommand{\thefigure}{\textbf{\arabic{figure}}}

\begin{document}

\title{Anomalous Fraunhofer patterns in gated Josephson junctions based on the bulk-insulating topological insulator BiSbTeSe$_{2}$}

\author{Subhamoy Ghatak\textsuperscript{\ddag}}
\author{Oliver Breunig\textsuperscript{\ddag}}
\author{Fan Yang\textsuperscript{\ddag}}
\email{yang@ph2.uni-koeln.de}
\author{Zhiwei Wang}
\author{A. A. Taskin}
\author{Yoichi Ando}
\email{ando@ph2.uni-koeln.de}
\affiliation{Physics Institute II, University of Cologne, Z\"ulpicher Str. 77, 
50937 K\"oln, Germany\\ 
{\rm (\textsuperscript{\ddag}These authors contributed equally to this work.)}}

\begin{abstract}
One-dimensional Majorana modes are predicated to form in Josephson junctions based on three-dimensional topological insulators (TIs). While observations of supercurrents in Josephson junctions made on bulk-insulating TI samples are recently reported, the Fraunhofer patters observed in such TI-based Josephson junctions, which sometimes present anomalous features, are still not well understood. Here we report our study of highly gate-tunable TI-based Josephson junctions made of one of the most bulk-insulating TI materials, BiSbTeSe$_{2}$, and Al. The Fermi level can be tuned by gating across the Dirac point, and the high transparency of the Al/BiSbTeSe$_{2}$ interface is evinced by a high characteristic voltage and multiple Andreev reflections with peak indices reaching 12. Anomalous Fraunhofer patterns with missing lobes were observed in the entire range of gate voltage. We found that, by employing an advanced fitting procedure to use the maximum entropy method in a Monte Carlo algorithm, the anomalous Fraunhofer patterns are explained as a result of inhomogeneous supercurrent distributions on the TI surface in the junction. Besides establishing a highly promising fabrication technology, this work clarifies one of the important open issues regarding TI-based Josephson junctions.
\end{abstract}

\maketitle

The superconducting proximity effect in the topological surface states of a three-dimensional (3D) topological insulator (TI) is of significant current interest, because it provides a promising means to engineer a two-dimensional (2D) topological superconducting state \cite{TI-JJ-theory-Fu,Sato-Ando}. Because of the spin-momentum locking \cite{TI-review-Hasan, TI-review-Qi, TI-review-Ando}, the spin degrees of freedom is frozen in the topological surface states and ``spinless" superconductivity is established when pairing is induced in surface-state electrons  proximitized by a conventional $s$-wave superconductor. The spinless nature is the key to realizing topological superconductivity associated with Majorana fermions \cite{Sato-Ando, MF-review-Elliott}, of which particle and hole excitations are identical. When Majorana fermions are spatially localized, they become Majorana zero-modes which obey non-Abelian statistics and are useful for topological quantum computation \cite{QuantumCpt-review-Sarma}. Hence, clarifying the nature of the proximity-induced superconductivity in the topological surface states of a 3D TI is not only interesting for finding topological superconductivity \cite{Sato-Ando} but also crucial for high-impact applications.

Josephson junctions (JJs) are the basic devices to investigate the superconducting proximity effect in 3D TIs. In a JJ, Andreev bound states formed in the weak link mediate dissipationless currents \cite{ABS-Klapwijk}. For a JJ based on a 3D TI, a pair of Majorana bound states, whose energy changes with the phase difference $\phi$ between the two superconductor (SC) electrodes with a $4\pi$ periodicity, are predicted to form  \cite{TI-JJ-theory-Fu, Snelder}. In the past, a variety of SC-TI-SC JJs \cite{TI-JJ-Samarth, TI-JJ-exp-Brinkman, TI-JJ-exp-LiLu, TI-JJ-exp-GoldhaberGordon, TI-JJ-exp-Lombardi, TI-JJ-exp-Juelich, TI-JJ-exp-Molenkamp, TI-JJ-exp-Mason, TI-JJ-exp-Lee, TI-JJ-exp-Brinkman2, Wiedenmann, TI-JJ-exp-Chen, Bocquillon} have been experimentally investigated to address these Majorana bound states. Although many of these experiments were carried out in JJs made of bulk-conducting materials \cite{TI-JJ-Samarth, TI-JJ-exp-Brinkman, TI-JJ-exp-LiLu, TI-JJ-exp-GoldhaberGordon, TI-JJ-exp-Lombardi, TI-JJ-exp-Juelich}, where a large part of the supercurrent is carried by conventional Andreev bound states, there have been several studies which were based on JJs made on bulk-insulating TIs \cite{TI-JJ-exp-Molenkamp, TI-JJ-exp-Mason, TI-JJ-exp-Lee, TI-JJ-exp-Brinkman2, Wiedenmann, TI-JJ-exp-Chen, Bocquillon}. However, the transparency of the interface between the SC and the TI surface states have not been so high compared to the case of InAs nanowires proximitized by epitaxially-grown Al \cite{Krogstrup}. Also, the experiments which employed electrostatic gating to approach the Dirac point have been relatively limited \cite{TI-JJ-exp-Molenkamp, TI-JJ-exp-Mason, Wiedenmann, TI-JJ-exp-Chen, Bocquillon}.

\begin{figure*}
\includegraphics[width=\linewidth]{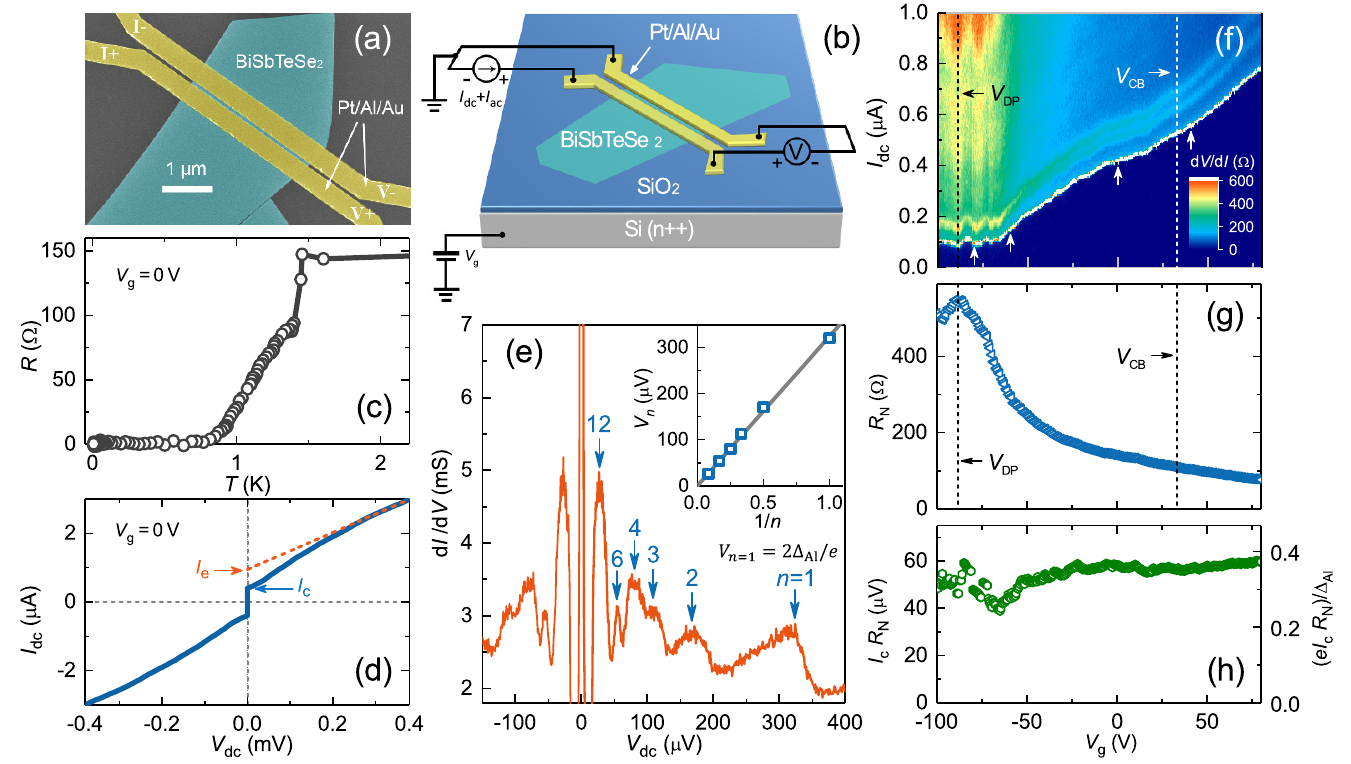}
\caption{\label{fig:fig1} (a) False-colour SEM image of device 1. Two 400-nm-wide Pt/Al/Au electrodes were fabricated with a distance of 70\,nm on a BiSbTeSe$_2$ flake, which was exfoliated on a Si/SiO$_2$ substrate, forming a Josephson junction with a width $W=2.9\,\mu$m.
(b) Schematics of the pseudo-four-wire measurement. Differential resistance $\textrm{d}V/\textrm{d}I$ of the junction was measured using a lock-in technique. The gate voltage $V_{\textrm{g}}$ was applied to the degenerately-doped Si substrate to tune the carrier density in the flake.
(c) Temperature dependence of the zero-bias resistance $R$, measured at $V_{\textrm{g}}=0\,$V, in device 1. (d) $I$-$V$ characteristic of device 1, measured at $T=8\,$mK and $V_{\textrm{g}}=0\,$V. Excess current $I_{\textrm{e}}$ is estimated by linearly extrapolating the high bias region to $V_{\textrm{dc}}=0\,$V. No hysteresis was observed between up- and down-sweeps.
(e) Plot of differential conductance $\textrm{d}I/\textrm{d}V$ vs $V_{\textrm{dc}}$ obtained in device 2. Sub-harmonic peaks of multiple Andreev reflections are indicated by arrows. Inset: Peak position $V_n$ vs inverse peak index $1/n$; a linear fit (solid line) gives $\Delta_{\textrm{Al}}=160\,\mu$eV.
(f) 2D mapping of $\textrm{d}V/\textrm{d}I$ as a function of $I_{\textrm{dc}}$ and $V_{\textrm{g}}$, measured at $T=8\,$mK. The white band at the edge of the superconducting region (deep blue) indicates the critical current $I_{\textrm{c}}$. The vertical arrows indicate the positions at which the Fraunhofer patterns in Figures 3a--3d were measured.
(g) Normal-state resistace $R_{\textrm{N}}$ as a function of $V_{\textrm{g}}$, measured at $T=8\,$mK and $B=60\,$mT, slightly above the critical field of the Al electrodes. The Dirac point at $V_{\textrm{DP}} \approx -88\,$V and the conduction band bottom at $V_{\textrm{CB}} \approx 33\,$V are indicated by vertical dashed lines in (f--g).
(h) $I_{\textrm{c}}R_{\textrm{N}}$ product (left axis) and $(eI_{\textrm{c}}R_{\textrm{N}})/\Delta_{\textrm{Al}}$ ratio (right axis) as functions of $V_{\textrm{g}}$. The data in (f--h) were acquired in device 1. }
\end{figure*}

Here, we present our study of highly gate-tunable Josephson junctions based on one of the most bulk-insulating TI materials, BiSbTeSe$_2$ \cite{RenBSTS}. A high characteristic voltage (expressed in the so-called ``$eI_{\textrm{c}}R_{\textrm{N}}/\Delta$ ratio" of $\sim$0.4, among the largest ever reported \cite{TI-JJ-exp-Brinkman, TI-JJ-exp-LiLu, TI-JJ-exp-GoldhaberGordon, TI-JJ-exp-Lombardi, TI-JJ-exp-Juelich, TI-JJ-exp-Molenkamp, TI-JJ-exp-Mason, TI-JJ-exp-Lee, TI-JJ-exp-Brinkman2, Wiedenmann, TI-JJ-exp-Chen}) demonstrates a high transparency of our SC/TI contact; this is also manifested in our observation of multiple Andreev reflections \cite{MAR-theory-Klapwijk} with the peak index reaching 12. By varying the gate voltage $V_{\text{g}}$, the Fermi level $E_{\text{F}}$ of BiSbTeSe$_2$ can be tuned in a broad range from inside the conduction band to below the Dirac point, establishing that proximity-induced surface superconductivity can be realized even at the Dirac point. Intriguingly, anomalous Fraunhofer patterns with missing or significantly suppressed lobes are observed in the entire range of $V_{\text{g}}$. Similar irregular Fraunhofer patterns have been reported in several cases \cite{TI-JJ-exp-Lee, Suominen, Bocquillon}, but the origin has not been clarified. In this regard, it was claimed that anomalous Fraunhofer patterns observed in TI-based JJs could be a signature of Majorana bound states \cite{TI-JJ-exp-GoldhaberGordon}. In the present work, we have been able to fit the anomalous Fraunhofer patterns by using a Monte Carlo algorithm involving the maximum entropy method and conclude that they result from an inhomogeneous supercurrent distribution.
In fact, it is understandable that the effect of current inhomogeneity becomes manifest in the surface-dominated transport regime where there is no bulk shunt.
This clarifies one of the important open issues regarding TI-based JJs, and our JJ fabrication technology paves the way for future Majorana devices in SC-TI hybrid systems.

The TI-based JJs reported here are fabricated on bulk-insulating BiSbTeSe$_{2}$ flakes.
A scanning electron microscope (SEM) image of a typical device is shown in Figure 1a as an example, and the device structure is schematically illustrated in Figure 1b. The flakes of BiSbTeSe$_{2}$ were exfoliated from a bulk single crystal and then transferred onto degenerately-doped Si substrates having a 290-nm-thick SiO$_2$ coating layer, which serves as the dielectric of the back gate. Flakes with a thickness $t$ of $\sim$20 nm or thinner were selected for device fabrication, so that the $E_F$ of the top surface can be controlled with the back gate. The JJ is defined by patterning two parallel Pt/Al/Au electrodes with a spacing $L$ to cover the full width $W$ of a selected flake; we call $L$ and $W$ the length and the width of the junction, respectively. Here the thin Pt layer works as a buffer layer to improve the transparency of Al/BiSbTeSe$_{2}$ contacts, and the Au capping layer helps to prevent Al electrodes from oxidization in air. We have investigated four JJ devices (denoted devices 1 -- 4) with essentially the same design (see Supporting Figure S1 for their pictures). The dimensions of all the devices are shown in Table S1 of the Supporting Information. The measurements were performed in a pseudo-four-wire configuration as illustrated in Figure 1b, by using both ac and dc techniques (see detailed descriptions at the end).

The temperature dependence of the junction resistance $R$ in device 1 is shown in Figure 1c. Superconducting transitions occur successively upon cooling: First, a  sharp drop in $R$ is observed at $T\sim1.4$ K, and then a broad transition continues down to 750 mK, below which the junction reaches a zero-resistance state. The first transition corresponds to the superconducting transition of Al electrodes, and the second transition is associated with the onset of Josephson coupling between the two Al electrodes.

A typical $I$-$V$ curve of device 1 measured at $V_{\textrm{g}}$ = 0 V is plotted in Figure 1d. For an applied dc current $I_{\textrm{dc}}$ smaller than the critical current $I_{\textrm{c}}=0.39\,\mu$A, the junction is in the superconducting state and the measured voltage $V_{\textrm{dc}}$ remains zero. When $I_{\textrm{dc}}$ exceeds $I_{\textrm{c}}$, a finite $V_{\textrm{dc}}$ starts to develop, and the junction is driven into a resistive state. However, due to quasiparticle processes such as multiple Andreev reflection, the $I$-$V$ curve is nonlinear at low $V_{\textrm{dc}}$. The linearity in the $I$-$V$ curve gradually recovers when $V_{\textrm{dc}}$ goes beyond $2\Delta_{\textrm{Al}}/e$; the inverse of the slope in this linear region gives a normal-state resistance $R_{\textrm{N}}^{I\textrm{-}V}=186\,\Omega$. An excess current $I_{\textrm{e}}=0.94\,\mu$A is obtained by linearly extrapolating the $I$-$V$ curve from the linear region to $V_{\textrm{dc}}=0\,$V, as illustrated in Figure 1c. Based on the Octavio-Tinkham-Blonder-Klapwijk theory \cite{OTBK-theory-Octavio}, the transparency $T_{\text{J}}$ of the Al/Pt/BiSbTeSe$_{2}$ contact can be estimated from the values of $R_{\textrm{N}}^{I\textrm{-}V}$ and $I_{\textrm{e}}$ \cite{OTBK-theory-Flensberg}, and $T_{\text{J}}=0.77$ is obtained for device 1. (see Supporting Information for details).

The $I$-$V$ curves of our JJs show no hysteretic behaviour; the values of $I_{\textrm{c}}$ obtained in up- and down-sweeps are identical. Such non-hysteretic $I$-$V$s are usually seen in overdamped JJs. The Stewart-McCumber parameter of device 1 is estimated to be $\beta_{c}=2eI_{\textrm{c}}(R_{\textrm{N}}^{I\textrm{-}V})^{2}C/\hbar \approx 1$, where $C$ is the shunt capacitance between the two Al electrodes. Similar values of $\beta_{c}$ are obtained for other devices (Supporting Information). Such intermediate values of $\beta_{c}$ indicate that our JJs lie in the region between the overdamped and underdamped limits. According to numerical calculations, JJs in such a region show non-hysteretic $I$-$V$ characteristics similar to those of underdamped JJs \cite{TI-JJ-exp-GoldhaberGordon, JJ-book-Likharev}, which is consistent with our observation.

Sub-harmonic peaks at $V_{n} = 2\Delta / ne$ are often observed in the $\text{d}I/\text{d}V$ vs $V_{\text{dc}}$ curves of voltage-biased JJs which satisfy $L<l_{\phi}$. Here, $2\Delta$ is the superconducting energy gap, $n=1,2...$ is the peak index and $l_{\phi}$ is the inelastic scattering length of the normal metal. Such sub-harmonic features are associated with multiple Andreev reflection (MAR) processes \cite{MAR-theory-Klapwijk} taking place in the channel of a JJ. In the past, the MAR peaks have been observed in a variety of TI-based JJs \cite{TI-JJ-Samarth, TI-JJ-exp-Molenkamp, TI-JJ-exp-Chen}.

In the present experiment, clear MAR peaks were seen in the $\text{d}I/\text{d}V$ vs $V_{\text{dc}}$ curves of all four devices. A typical curve measured in device 2 is plotted in Figure 1e with assigned peak indices. The assignments of $n$ are confirmed by the linear relation between  $V_{n}$ and $1/n$ illustrated in the inset of Figure 1e.  A linear fit to the $V_{n}(1/n)$ data gives $\Delta_{\text{Al}}=160\,\mu$eV, consistent with the superconducting gap $\Delta_{\text{Al}}$ reported for Al in the literature \cite{TI-JJ-exp-GoldhaberGordon}.

A key feature in Figure 1e is the presence of high-index conductance peaks reaching $n$ = 12, indicating that the inelastic scattering length $l_\phi$ in BiSbTeSe$_{2}$ exceeds $12L$ (= 840 nm), which is longer than what was reported for previous TI-based JJs \cite{TI-JJ-exp-GoldhaberGordon}. Nevertheless, such a long $l_\phi$ is consistent with previous results of Aharonov-Bohm oscillations observed in square-ring interferometers based on  Bi$_2$Se$_3$ flakes \cite{AB-effect-LiLu}. In addition, the presence of high-index peaks also indicates a relatively high transparency $T_{\text{J}}$ of the Al/BiSbTeSe$_{2}$ contacts. Numerical calculations \cite{MAR-peak-index-Yeyati, MAR-peak-index-Bardas} show that high-index features of MAR become the most visible when $T_{\text{J}}$ takes a relatively high value of around $0.7-0.8$, while they disappear in both transparent ($T_{\text{J}}=1$) and tunneling ($T_{\text{J}}=0$) limits, which is consistent with our estimate of $T_{\text{J}}$ = 0.83 for device 2.

The amplitude of the conductance peak at $V_n$ is expected to decay with increasing $n$, because the probability of MAR processes generally decreases with $n$. However, in our experiment, the $n=12$ peak shows an amplitude much larger than those of other peaks. The origin of the larger-than-expected amplitude of $n=12$ peak is not clear and the reason for the absence of peaks for $n=7-11$ is unknown. Nevertheless, similar MAR phenomena have been seen in previous experiments on TI-based JJs by Jauregui et al. \cite{TI-JJ-exp-Chen} and on Ge/Si-nanowire-based JJs by Xiang et al. \cite{Xiang2006}.

As-grown BiSbTeSe$_{2}$ single crystals are bulk-insulating, with $E_{\text{F}}$ located slightly below the Dirac point \cite{BSTS2-sato}. However, due to the unintentional doping introduced during the exfoliation and device fabrication processes, $E_{\text{F}}$ in our samples is shifted into the $n$-type region. Nevertheless, the top-surface Fermi level of our devices can be tuned by the back-gate voltage $V_{\text{g}}$ to the Dirac point, thanks to the small thickness ($\sim$18 nm) of the TI flakes used here; in the following, $V_{\text{DP}}$ denotes the $V_{\text{g}}$ value needed to bring the Fermi level to the Dirac point. For a detailed analysis of the single-gate control of the Fermi levels of two surfaces involving a capacitive coupling, see Ref. \citenum{Yang2015}.

The normal-state resistance $R_{\text{N}}$ vs $V_{\text{g}}$ curve of device 1 is plotted in Figure 1g. Here, $R_{\text{N}}$ was measured by suppressing the superconductivity of Al with a magnetic field; note that $R_{\text{N}}$ and $R_{\text{N}}^{I\textrm{-}V}$ are not exactly the same, which probably indicates that the contact resistance is very sensitive to the magnetic field. The position of the Dirac point $V_{\text{DP}}=-88\,$V is determined by the peak in this curve, while the position of the conduction band bottom $V_{\text{CB}}$ is estimated by using $V_{\text{CB}}=en_{2d}/C_{\text{g}}+V_{\text{DP}}\approx 33\,$V, where $n_{2d}\approx 9.1 \times 10^{12}\,\text{cm}^{-2}$ is the two-dimensional carrier density in BiSbTeSe$_{2}$ when $E_{\textrm{F}}$ reaches the bottom of the conduction band, and $C_{\text{g}}\approx12\,\text{nF/cm}^2$ is the capacitance per unit area of the back gate (see Supporting Information for details). 

Figure 1f shows the colour mapping of $\text{d}V/\text{d}I$ in device 1 scanned in the $I_{\rm dc}$ vs $V_{\rm g}$ plane. The critical current $I_{\text{c}}$ is depicted by the white band at the edge of the superconducting region in deep blue colour. A finite $I_{\text{c}}$ of 92 nA is observed at the Dirac point. In the $n$-type region where $V_{\text{g}} > V_{\text{DP}}$, $I_{\text{c}}$ increases sharply with increasing $V_{\text{g}}$. However, in the $p$-type region where $V_{\text{g}}<V_{\text{DP}}$, $I_{\text{c}}$ saturates when $V_{\text{g}}$ decreases. In device 2 where $E_{\text{F}}$ can be tuned deeper into the $p$-type region, we found that $I_{\text{c}}$ actually increases slowly with decreasing $V_{\text{g}}$ in the $p$-type region, although the rate of increase is very low (see Supporting Figure S2c). In the past, similar asymmetric $I_{\text{c}}(V_{\text{g}})$ behavior was also reported in TI-based JJs \cite{TI-JJ-exp-Mason, TI-JJ-exp-Chen} and graphene \cite{Graphine-JJ-exp-Geim}, and it was speculated that this behaviour is caused by the $p$-$n$ junctions formed between the gated junction channel and the areas under the SC contacts.

The $I_{\text{c}}R_{\text{N}}$ product of device 1 is calculated as a function of $V_{\text{g}}$ using the data in Figures 1f and 1g, and shown in Figure 1h. The $I_{\text{c}}R_{\text{N}}$ product is almost independent of $V_{\text{g}}$ except for the region near the Dirac point. When $E_{\text{F}}$ approaches the Dirac point from the $n$-type region, $I_{\text{c}}R_{\text{N}}$ presents a minimum at around $V_{\text{g}}-V_{\text{DP}}\approx35\,$V.
Similar behavior in $I_{\text{c}}R_{\text{N}}$ was observed with less clarity in device 2
(see Supporting Figure S2d). The origin of this minimum is not clear. Most importantly, device 1 shows the $eI_{\textrm{c}}R_{\textrm{N}}/\Delta$ ratio of up to 0.37, which is among the largest values reported in the literature \cite{TI-JJ-exp-Brinkman, TI-JJ-exp-LiLu, TI-JJ-exp-GoldhaberGordon, TI-JJ-exp-Molenkamp, TI-JJ-exp-Mason, TI-JJ-exp-Lee, TI-JJ-exp-Lombardi, TI-JJ-exp-Brinkman2, TI-JJ-exp-Juelich, TI-JJ-exp-Chen}, again proving the high transparency of the Al/BiSbTeSe$_2$ contacts realized in the present work.

\begin{figure}
\includegraphics[width=0.8\linewidth]{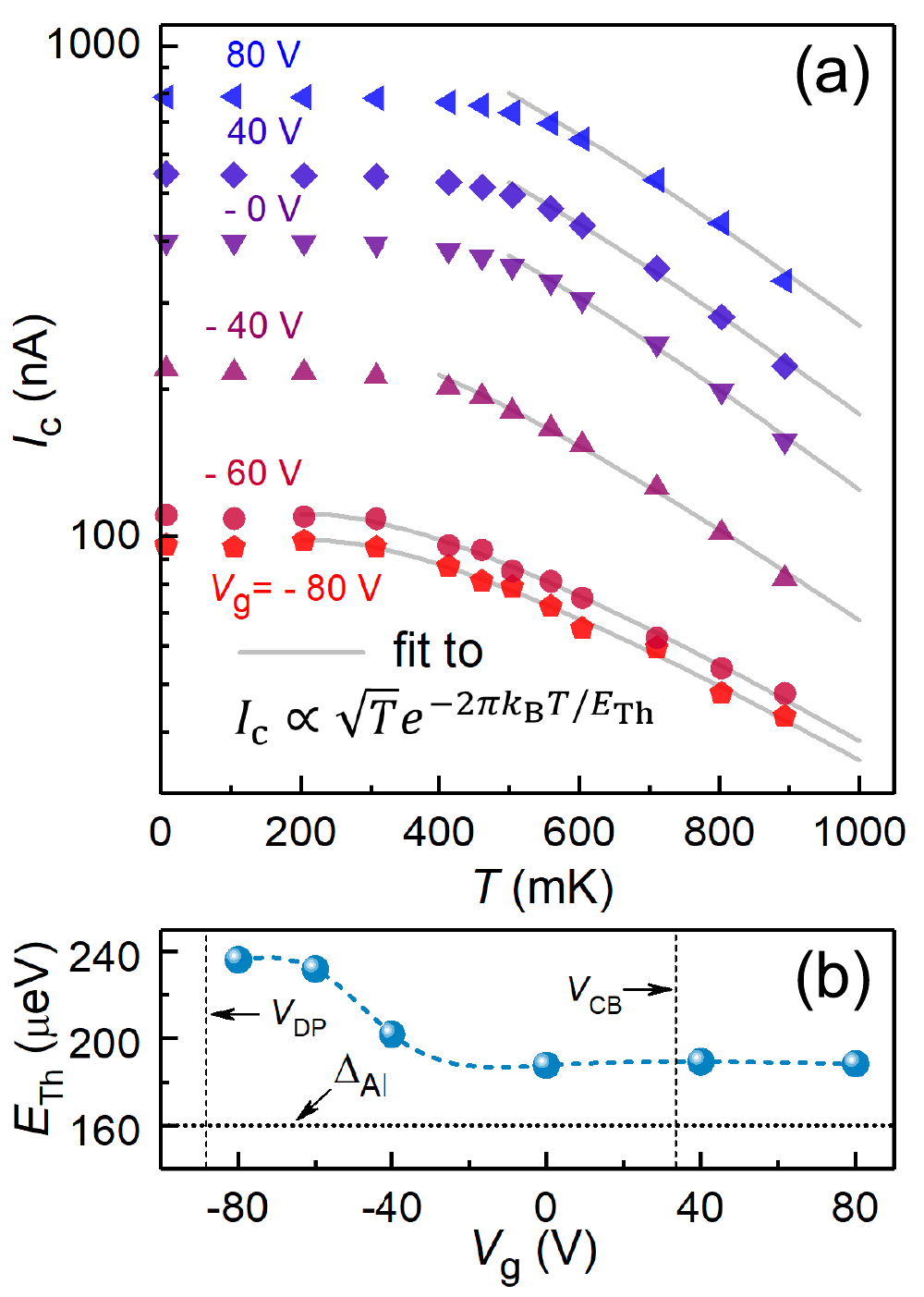}
\caption{\label{fig:fig2} {(a) Temperature dependences of the critical current $I_{\textrm{c}}$ at various $V_{\textrm{g}}$, measured in device 1. The values of $I_{\textrm{c}}$ were determined from the maxima in the $\textrm{d}V/\textrm{d}I$ vs $I_{\textrm{dc}}$ curves. The gray lines are the best fits to $I_{\textrm{c}} \propto \sqrt{T}\,\textrm{exp} \left(-2\pi k_{\textrm{B}}T/E_{\textrm{Th}}\right)$.
(b) Gate-voltage dependence of the Thouless energy $E_{\textrm{Th}}$ obtained from the fitting. The positions of the Dirac point and the conduction band bottom are shown by vertical dashed lines. Horizontal dotted line shows the value of $\Delta_{\textrm{Al}}$.}}
\end{figure}

The temperature dependence of $I_{\text{c}}$ was measured at various $V_{\text{g}}$ values, as plotted in Figure 2a. With decreasing $T$, $I_{\text{c}}$ first increases exponentially and then saturates. The step-like increase in $I_{\text{c}}$ at low temperature reported in a recent work on TI-based JJs using Nb electrodes \cite{TI-JJ-exp-Juelich} was not observed in our devices. A fit to $I_{\textrm{c}} \propto\sqrt{T}\,\exp \left(-2\pi k_{\textrm{B}}T /E_{\textrm{Th}} \right)$ in the high-temperature range allows us to extract the Thouless energy $E_{\textrm{Th}}$ ($\equiv \hbar D/L^2$) at various $V_{\text{g}}$ \cite{TI-JJ-exp-Lombardi,CPR-review-Golubov}; here, $D$ is the average diffusion constant of supercurrent-carrying electrons and $L$ is the length of the JJ.

The $E_{\textrm{Th}}$ values obtained from the fittings are plotted in Figure 2b as a function of $V_{\text{g}}$. It is noteworthy that $E_{\textrm{Th}}$ remains unchanged when $E_{\textrm{F}}$ moves through the bottom of the conduction band at $V_{\text{g}}=V_{\text{CB}}$. In other words, the presence of bulk carriers does not change the average $D$ of the supercurrent-carrying electrons. Since $D$ of bulk carriers is expected to be much smaller than that of the surface electrons, the unchanged $D$ suggests that the supercurrent is mainly carried by the surfaces states even when $E_{\textrm{F}}$ is located inside the conduction band. This is in agreement with the conclusions in previous studies on planar TI-based JJs \cite{TI-JJ-exp-Brinkman, TI-JJ-exp-Mason}, but contradicts the results obtained in Pb-Bi$_2$Se$_3$-Pb JJs fabricated along the $c$-direction of Bi$_2$Se$_3$ flakes \cite{TI-JJ-exp-LiLu}. As $E_{\textrm{F}}$ approaches the Dirac point from the $n$-type region, $E_{\textrm{Th}}$ starts to increase, indicating an increase in $D$ of the surface electrons. Such an enhancement of $D$ around the Dirac point is probably due to the smaller phase space available for scattering. In addition, the extracted $E_{\textrm{Th}}$ is comparable to $\Delta_{\text{Al}}$ in the entire range of $V_{\text{g}}$, indicating that our junctions lie on the boundary between the short- and long-junction limit, similar to what was previously reported in Al-Bi$_2$Se$_3$-Al junctions\cite{TI-JJ-exp-Lombardi}.

To determine the regime to which our JJs belong, it is important to estimate several characteristic lengths.
As discussed above, the lower boundary of the inelastic scattering length $l_{\phi}$ is estimated from the MAR results to be $l_{\phi} > 12L$ = 840 nm.
The superconducting coherence length $\xi_{\text{N}}$ in BiSbTeSe$_{2}$ is given by $\xi_{\text{N}}=\sqrt{\hbar D/\Delta_{\text{Al}}}=\sqrt{E_{\text{Th}}L^2/\Delta_{\text{Al}}}$. With $L$ = 70 nm measured in the SEM image, $\Delta_{\text{Al}}=160\,\mu$eV obtained from the MAR data, and $E_{\text{Th}}$ = 187 -- 236 $\mu$eV extracted from the $I_{\text{c}}(T)$ curves, we obtain $\xi_{\text{N}}=76-85\,$nm.
The mean free path $l_{\text{e}}$ of the supercurrent-carrying surface electrons can be calculated by using $E_{\text{Th}}=\hbar D/L^2$ and $D=v_{\text{F}}l_{\text{e}}/2$, where $v_{\text{F}}=5.5\times10^5\,$m/s is the Fermi velocity of the surface states \cite{BSTS2-sato}, and one obtains $l_{\text{e}}$ = 5.1 -- 6.4 nm, which is much shorter than the previously reported value $l_{\text{e}}\sim40\,$nm in Bi$_{1.5}$Sb$_{0.5}$Te$_{1.7}$Se$_{1.3}$ bulk crystals \cite{BSTS-transport-Ando}. We tentatively attribute the short $l_{\text{e}}$ in the BiSbTeSe$_{2}$ flakes to the surface disorder introduced in the device fabrication process. While a short $l_{\text{e}}$ might seem a disadvantage for quantum computing, the readout of Majorana qubits can be performed without relying on ballistic transport \cite{Schrade}. 
The characteristic lengths estimated above point to the relation $l_{\text{e}}\ll L \sim \xi_{\text{N}} \ll l_{\phi}$, which indicates that our JJs are diffusive junctions ($L\gg l_{\text{e}}$) in the intermediate region between the short- and long-junction limit ($L \sim \xi_{\text{N}}$).

\begin{figure*}
\includegraphics[width=\linewidth]{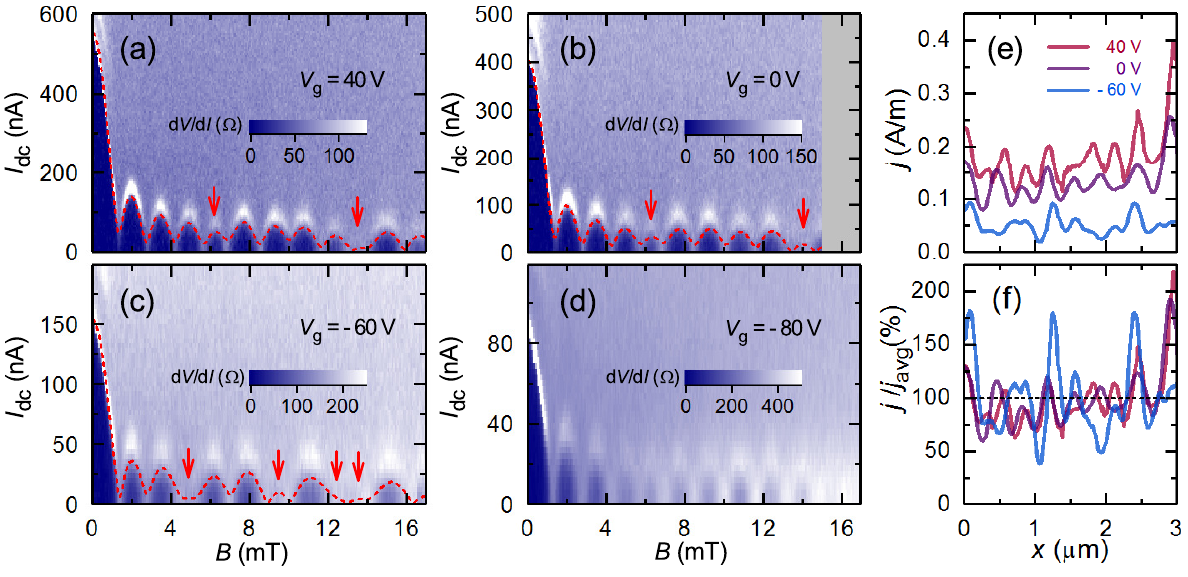}
\caption{\label{fig:fig3} {(a-d) 2D plots of $\textrm{d}V/\textrm{d}I$ as a function of $I_{\textrm{dc}}$ and $B$ at fixed $V_{\textrm{g}}$ values of (a) $40\,$V, (b) $0\,$V, (c) $-60\,$V and (d) $-80\,$V, measured at $T=8$\,mK in device 1. Irregular Fraunhofer patterns were observed at all gate voltages. Red arrows indicate the missing or suppressed lobes. The anomalous Fraunhofer patterns are understood in terms of inhomogeneous supercurrent distributions. Red dashed lines represent the
fits to the experimental data obtained by an advanced fitting scheme employing the maximum entropy method in a Monte Carlo algorithm. The pattern at $V_{\textrm{g}}=-80\,$V was not fitted, because the $I_{\textrm{c}}$ values could not be reliably extracted.
(e-f), The inhomogeneous current density distribution $j(x)$ (e) and its normalized plot $j(x)/j_{\textrm{avg}}$ (f) obtained as a result of the fittings that yielded the red dashed lines in (a-c).
}}
\end{figure*}

For JJs satisfying $L \leq \xi_{\text{N}}$, the $eI_\text{c}R_\text{N}/\Delta$ ratio is expected to be of order 1 at $T=0\,$K \cite{JJ-review-Likharev}. However, in reality, the measured values of $eI_\text{c}R_\text{N}/\Delta$ are usually much smaller than the theoretical expectation due to the existence of resistive shunt channels or limited SC contact transparency \cite{TI-JJ-exp-Lombardi}. The $eI_\text{c}R_\text{N}/\Delta$ ratio found in previous studies on TI-based JJs ranges between 0.01 and 0.31, all smaller than the ratio of 0.37 observed in device 1, indicating that the quality of our TI-based JJs is among the best in terms of bulk-insulating property and contact transparency.

For a trivial JJ with a homogeneous current distribution, the $I_{\text{c}}$ vs $B$ relation is expected to exhibit a so-called Fraunhofer pattern described by
\begin{equation}
I_{\text{c}}(B)=I_{\text{c0}}\left|\frac{\mathrm{sin}(\pi\Phi_{\text{J}}/\Phi_{0})}{\pi\Phi_{\text{J}}/\Phi_{0}}\right|,
\end{equation}
where $I_{\text{c0}}$ is the critical current at $B=0$, $\Phi_{\text{J}}$ is the flux threading through the junction channel and $\Phi_{\text{0}}=h/(2e)$ is the flux quantum.
To elucidate the $I_{\text{c}}(B)$ behaviour of our JJs, the $\text{d}V/\text{d}I$ data were measured as a function of $I_{\text{dc}}$ and $B$ at various $V_{\text{g}}$ corresponding to the $E_{\text{F}}$ positions from inside the conduction band to near the Dirac point. The $I_{\text{c}}(B)$ patterns of device 1 are shown in Figures 3a--3d. All the observed patterns are essentially Fraunhofer-like, but bear anomalous features that deviate from a standard Fraunhofer pattern. They include: (i) an average period $\Delta B_{\text{avg}}\approx 1.44\,$mT, which is about three times smaller than the expected period $\Delta B_{0}\approx 5.29\,$mT estimated from the junction area $S=(L+2\lambda_{\text{Al}})W$, where $\lambda_{\text{Al}}$ ($\approx 50\,$nm) is the London penetration depth of bulk Al, and (ii) missing or significantly suppressed lobes which cannot be described by the formula for the Fraunhofer pattern.

Anomalous features in the Fraunhofer patterns have been reported for TI-based JJs in some of the previous works \cite{TI-JJ-exp-GoldhaberGordon,TI-JJ-exp-Lee, Suominen, Bocquillon}. There was an attempt to explain the smaller-than-expected $\Delta B_{\text{avg}}$ by a phenomenological model involving quantized Majorana modes \cite{TI-JJ-exp-GoldhaberGordon}. However, in the present case, it may partly be explained by a small 2D superfluid density (and hence longer effective $\lambda_{\text{Al}}$) in the 30-nm-thick Al electrodes which are thinner than the bulk $\lambda_{\text{Al}}$.
Also, it has been noted that the small $\Delta B_\text{avg}$ can be explained by the flux focusing effect \cite{TI-JJ-exp-LiLu}, which is commonly seen in planar JJs \cite{flux-focusing-Rosenthal}. Due to the diamagnetism of a SC, the magnetic field near the SC electrodes is distorted, and half of the magnetic flux expelled from the SC electrodes is compressed into the junction; this effect may be quantified in terms of an extended effective junction area $S_\text{eff}=WL_\text{eff}$, where $L_\text{eff}=(L+\frac{d_1}{2}+\frac{d_2}{2})$ is the effective junction length considering the flux focusing ($d_1$ and $d_2$ are the widths of the left and right SC electrodes). Taking $d_1 = d_2 =400\,$nm and $L=70\,$nm for device 1, we obtain $\Delta B=\Phi_0/(S_\text{eff})=1.52\,$mT, which is close to the observed average period $\Delta B_{\text{avg}}\approx 1.44\,$mT. Therefore, in the following, we use $S_\text{eff}$ instead of $S$ for the reconstruction of the $I_\text{c}(B)$ patterns.

Regarding the origin of the other feature, \textit{i.e.} missing lobes in the Fraunhofer pattern, this has been left unexplained in the literature \cite{TI-JJ-exp-Lee, Suominen, Bocquillon}.
Theoretically, an irregular $I_\text{c}(B)$ pattern could be generated by (i) a non-sinusoidal current phase relation or (ii) an inhomogeneous supercurrent distribution in the JJ. In the past, much attention has been paid to the first possibility. For instance, it was predicted that the formation of 1D Majorana modes in a TI-based JJ gives rise to an $I_\text{c}(B)$ pattern featuring a non-zero first minimum, deviating from a standard Fraunhofer pattern \cite{LiangFu-2013}. It was also predicted that the $4\pi$ periodicity due to Majorana zero-modes leads to a suppression of the odd-numbered (1st, 3rd, 5th, etc.) lobes in the Fraunhofer patttern \cite{Baxevanis}.

However, these theoretical possibilities do not really explain the disappearance of some particular lobes, and one had better consider the second possibility to understand the irregular $I_\text{c}(B)$ patterns observed in our TI-based JJs. We set out to reconstruct the anomalous Fraunhofer patterns by considering an inhomogeneous supercurrent distribution $j(x)$ along the width of the junction. The experimental data of $I_\text{c}(B)$, extracted from the $\frac{\mathrm{d}V}{\mathrm{d}I}(I_\mathrm{dc})$ data as the onset of rapid increase, are fitted by using a Monte Carlo algorithm to optimize the form of $j(x)$ based on the maximum entropy method (see detailed descriptions at the end), which has not been applied to the present problem; this method removes the constraint of the standard Dynes-Fulton-type analysis \cite{Dynes1971,Allen2016,Zappe1975,Nesher1997,Hui2014} and allows to identify any form of inhomogeneous supercurrent distribution.
In Figure 3e we plot for three $V_{\rm g}$ values the obtained forms of $j(x)$, which reproduce all the main features in the $I_\text{c}(B)$ patterns as shown with red dashed lines in Figures 3a--3c. In the normalized plot of $j(x)/j_{\rm avg}$ shown in Figure 3f, one can see that the relative magnitude of the fluctuations is enhanced at $V_{\rm g} = -60$ V, which is close to the Dirac point. The enhancement of $j(x)$ at the edges seen in Figure 3f is probably due to the contribution of the side surfaces \cite{TI-JJ-exp-Lee} which also contain topological surface states.
We did not perform a fitting of the $I_\text{c}(B)$ data for $V_{\rm g} = -80$ V, because the $I_{\textrm{c}}$ values at this $V_{\rm g}$ could not be reliably extracted due to the absence of true supercurrents in magnetic fields of only a few mT.

We note that the $j(x)$ curves presented in Figure 3e are not the unique solutions which give good fittings to the data. This is partly because of the lack of the phase information in this inverse problem, but also because of the trade-off between the degree of fitting and the smoothness of the solution. Nevertheless, it is clear from the present analysis that the existence of an inhomogeneous $j(x)$ is sufficient for causing the observed anomalous Fraunhofer patterns, most notably missing lobes.

An inhomogeneous $j(x)$ can be caused by various reasons. One possible reason is the fluctuations in the junction length $L(x)$ due to the roughness of the edges of Al electrodes. Since the critical current depends exponentially on $L$, for junctions with a short $L$, even small fluctuations in $L(x)$ could result in a notable variation in $j(x)$. Another possible reason is a position-dependent transparency of the Al/TI interface, which may arise from a granularity of the Al film. In addition, a possible variation of the surface carrier density due to charge puddles \cite{puddles-Breunig2017} may also be partially responsible for the inhomogeneous $j(x)$.

To conclude, we have established a highly promising fabrication technology for TI-based JJs, which allowed us to elucidate that the anomalous Fraunhofer patterns are essentially due to an inhomogeneous supercurrent distribution in the proximitized surface states. This technology and the quantitative understanding gained in this work would make it possible to fabricate an optimized device to nail down the signature of Majorana fermions in the proximitized TI surface in near future.

\vspace{5mm}

{\bf Growth of bulk BiSbTeSe$_{2}$ crystals:}
The bulk BiSbTeSe$_{2}$ single crystals were grown by melting high-purity (99.9999\%) Bi, Sb, Te, and Se shots with the
molar ratio of 1:1:1:2 in an evacuated quartz tube via a modified Bridgman method, following our established recipe reported in Ref. \citenum{RenBSTS}. The grown crystals typically have a domain size of several mm and are easily cleaved along the (111) plane.

{\bf Device fabrication:}
Ultrathin BiSbTeSe$_{2}$ flakes are obtained by mechanical exfoliation of bulk single crystals on degenerately-doped Si/SiO$_2$ wafers by a Scotch-tape technique. Promising flakes are initially identified under a laser microscope (Keyence VK-X200), and further characterized by atomic force microscopy. The flakes with an extremely smooth surface (without atomic terraces) are chosen for the fabrication of JJ devices using electron beam lithography. Previous studies on a similar bulk-insulating TI material have shown \cite{TI-JJ-exp-Brinkman2} that a very short junction channel ($\sim$50 nm) is essential for realizing a sufficiently strong coupling between two SC electrodes to observe a Josephson supercurrent. To reliably fabricate short-channel junctions, we have developed our own recipe. A 160-nm-thick PMMA  layer was spin coated and baked at 120$^{\circ}$C for 10 minutes. The reduced baking temperature ensures minimal unintentional doping due to creation of Se vacancies at elevated temperatures. The resist layer was exposed by a 25-kV electron beam. The width of the Al electrodes was kept around 400 nm to avoid vortex movements at finite magnetic fields in the superconducting state. To remove any resist residues in the contact areas after the development, the wafers were cleaned with oxygen plasma in a reactive-ion etching machine at 20 W for 7 s. Afterwards, the wafers were further cleaned in a HCl solution to remove any oxide layer in the contact areas. The metallization was carried out by sputtering 5-nm-thick Pt as buffer layer, followed by thermal evaporation of 30-nm-thick Al at the pressure of $5\times10^{-7}$~mbar. A 5-nm-thick Au layer was successively evaporated to protect the Al from oxidation, and the lift-off was done in warmed acetone.

{\bf Measurements:}
The devices were cooled in a dry dilution refrigerator (Oxford Instruments TRITON 200) with a base temperature of less than 10 mK. To reduce the electromagnetic noise, the electrical lines are equipped with RC-filters at room temperature and at the 4-K plate, along with additional RC and copper-powder filters at the mixing chamber plate. The junctions were electrically characterized by dc current-voltage ($I$-$V$) and differential resistance (${\rm d}V/{\rm d}I$) measurements in a pseudo-four-probe configuration. For the ${\rm d}V/{\rm d}I$ measurement, an 18.6-Hz ac current with the amplitude of 1 nA was superimposed to a dc bias current. The magnetic-field dependence of the critical current (\textit{i.e.} Fraunhofer pattern) was measured by performing the ${\rm d}V/{\rm d}I$ measurements while driving the superconducting magnet with a Keithley 2450 source meter.

{\bf Current density retrieval:}
For a JJ with a sinusoidal current-phase relation, the experimentally observed critical current $I_c$ is related to the modulus of the phase-dependent integration of the supercurrent density \cite{Dynes1971},
\begin{align} \label{eqn:IcFT}
I_c^\mathrm{FT}(B)= \left| \int\limits_{-W/2}^{W/2} j(x) e^{i \varphi(x,B)} \,dx \right|  ,
\end{align}
which allows to deduce information about the distribution of the current density $j$ and the phase difference $\varphi$ within the junction.
In the present case of a narrow JJ (\textit{i.e.} $W \ll \lambda_{\text{J}}$, where $\lambda_{\text{J}}$ is the Josephson penetration length), $\varphi$ can be assumed to depend linearly on the spatial position $x$ along the junction and on the applied magnetic field $B$; namely, $\varphi(x,B)=2\pi x L_\mathrm{eff} B / \Phi_0  + \varphi_0$, where $\Phi_0=h/2e$ is the magnetic flux quantum and $L_\mathrm{eff}=L+2 \cdot \frac{d}{2}$ is the effective length of the JJ to take into account the flux focusing effect of the superconducting electrode of width $d$ \cite{flux-focusing-Rosenthal}.
This phase factor makes the integration to be effectively a Fourier transform from the $x$ coordinate to the $B$ coordinate. In general, the supercurrent density $J(x,y)$ may depend on both the spatial coordinates $x$ and $y$ within the junction area; however, we restrict ourselves to the dependence on $x$ defined along the junction's width $W$, by considering the one-dimensional current density $j(x)=\int\limits_{-L/2}^{L/2} J(x,y)\,dy$.

As the critical current does not contain information on the complex phase of the Fourier transform, but only on its modulus, the current density $j(x)$ cannot be obtained unambiguously from the experimental data $I_c^\mathrm{exp}(B)$. In fact, there is an infinitely large number of current profiles $j(x)$ that yield a similar $I_c^\mathrm{FT}(B)$.
Most of previous reconstructions of the current distribution were essentially based on the Dynes-Fulton analysis \cite{Dynes1971,Allen2016,Zappe1975,Nesher1997,Hui2014}. Although the original Dynes-Fulton analysis utilized a relatively strong assumption that $j(x)$ is an approximately even function \cite{Dynes1971}, some of the later attempts have relaxed this constraint \cite{Zappe1975,Nesher1997,Hui2014}. 
In this work, we perform a novel type of analysis based on the ideas of Dynes and Fulton, but without requiring any presumptions; specifically, we identify the most physical solution by using a maximum entropy method \cite{Jaynes}.
The outline of our fitting procedure is the following:
First, we discretize $j(x)$ into $N$ elements $j_i$ and initialize them to a value of $I_c^\mathrm{exp}(0)/N$. To ensure the agreement with the zero-field critical current, we keep $\sum\limits_i j_i=I_c^\mathrm{exp}(0)$ to be fixed. Then we fit the data by maximizing the functional $\mathcal{F}[j]= -\lambda \chi_j^2-\sum\limits_i \tilde{j}_i \ln \tilde{j}_i$ using a Monte Carlo algorithm.
Here, $\chi_j^2=\sum\limits_{k=0}^{M} \left( I_c^\mathrm{FT}(B_k)-I_c^\mathrm{exp}(B_k)\right)^2 $ is the square of the deviation from the experimental data taken at points $B_k$, and the second term is the entropy of the normalized current distribution $\tilde j_i=j_i/\sum\limits_i j_i$.
The weighting factor $\lambda$ tunes between a pure least-square fit for $\lambda\rightarrow\infty$ and an increasing contribution of the maximum entropy method for $\lambda\rightarrow 0$. A larger $\lambda$ gives a better fitting, but the obtained current distribution tends to become unphysically peaky; more weighting of the entropy term makes the resulting current distribution to be smoother and more physical, even though $\chi_j^2$ becomes larger.
In our analysis, the value of $\lambda$ is initially set to $1/(2M)$, and we increase (decrease) it during the Monte Carlo steps when the iteration goes to the wrong direction, \textit{i.e.}, when $\chi_j^2$ becomes larger (smaller) than a given target value $\xi$; such a dynamical tuning of $\lambda$ allows the square term $\chi_j^2$ to converge to $\xi$ while ensuring the resulting current distribution to be smooth.
Still, the algorithm does not provide a single and unique solution for the current distribution, and the smoothness of the distribution depends on the choice of $\xi$. To avoid over-fitting of the data, for each data set the maximal $\xi$ was chosen so that the calculated  $I_c^\mathrm{FT}(B)$ reproduces the absent lobes and the overall shape of the Fraunhofer pattern. Typically, the fits converged after about 50,000 steps. Each data set was processed about 100 times to check for the robustness of the fitting result, and we found that the obtained current distributions -- save for mirror symmetry -- resembled each other within a few percent of deviation. From these we pick the solution having the largest entropy and show it in Figure 3.

{\bf Author contributions:}
S.G., O.B. and F.Y contributed equally to this work and should be considered as co-first authors. 

{\bf Acknowledgement:}
This project has received funding from the European Research Council (ERC) under the European Union's Horizon 2020 research and innovation programme (grant agreement No 741121) and was also supported by DFG (CRC1238 ``Control and Dynamics of Quantum Materials", Projects A04 and B01). We thank D. Goldhaber-Gordon, A. Rosch, S. Trebst, and Y. Vinkler for helpful discussions.
O.B. acknowledges the support from Quantum Matter and Materials Program at the University of Cologne funded by the German Excellence Initiative.


\clearpage
\onecolumngrid

\renewcommand{\thefigure}{S\arabic{figure}} 
\renewcommand{\thetable}{S\arabic{table}}

\setcounter{figure}{0}

\begin{flushleft} 
{\Large {\bf Supporting Information}}
\end{flushleft} 
\vspace{2mm}

\begin{flushleft} 
{\bf 1. SEM images of all devices}
\end{flushleft}

The SEM images of all four devices are shown in Figure S1.

\begin{figure}[h]
\centering
\includegraphics[width=1.0\textwidth]{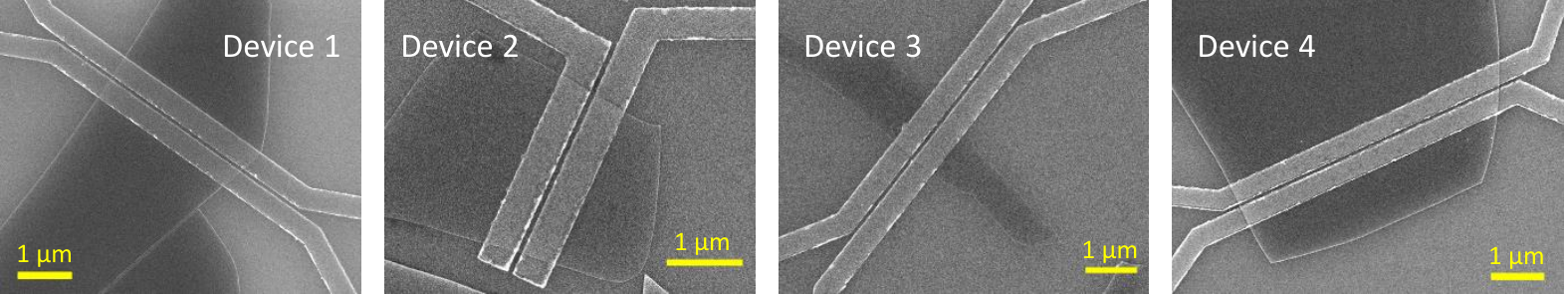}
\caption{Scanning electron microscope (SEM) images of all four devices. The scale bars are 1~$\mu$m. }
\end{figure}

\begin{flushleft} 
{\bf 2. Characteristic parameters of all devices}
\end{flushleft}

Characteristic parameters of all four devices are listed in Table S1. Here, $R_{\text{N}}$ denotes the normal-state resistance measured by applying a magnetic field $B>B^\text{Al}_\text{c}$ or a large excitation current $I_\text{ac}\gg I_\text{c}$, while $R_{\text{N}}^{I\textrm{-}V}$ is the normal-state resistance extracted from $I$-$V$ curves. Since $R_{\text{N}}$ is obtained with Al in the normal-conducting state one may naively expect that $R_\text{N}>R_{\text{N}}^{I\textrm{-}V}$. Yet, the opposite is observed, which may be explained by a change of the density of states in the TI due to the proximity effect \cite{Gueron-2012}. 

\begin{table}[h]
\setlength{\tabcolsep}{1pt}
\footnotesize
\renewcommand{\arraystretch}{1.1}
\caption{\label{table1}
Characteristic parameters of all devices at $V_{\textrm{g}}$ = 0 V (calculated with $\Delta_{\text{Al}}$ = 160 $\mu$eV).}
\begin{tabular}{|c|c|c|c|c|c|c|c|c|c|c|c|c|c|} \hline
 No. & \,$L$~(nm)\,& \,$W$~($\mu$m)\, & \,$t$~(nm)\, & \,$I_{\text{c}}$~($\mu$A)\, & \,$I_{\text{e}}$~($\mu$A)\, & \,$R_{\text{N}}$~($\Omega$)\, & \,$R_{\text{N}}^{I\textrm{-}V}$~($\Omega$)\, & \,$I_{\text{c}}R_{\text{N}}$~($\mu$V)\, & \,$ I_{\text{c}}R_{\text{N}}^{I\textrm{-}V}$($\mu$V)\, & \,$eI_{\text{c}}R_{\text{N}}/\Delta_{\text{Al}}$\, & \,$eI_{\text{c}}R_{\text{N}}^{I\textrm{-}V}/\Delta_{\text{Al}}$\, & \,\,$\beta_{\text{c}}$\,\, & \,\, $T_{\text{J}}$\,\, \\  \hline
 1 & 70 & 2.9 & 18 & 0.39 & 0.94 & 141 & 187 & 55 & 73 & 0.34 & 0.45 & 0.99 & 0.77 \\  \hline
 2 & 70 & 2.2 & 17 & 0.15 & 0.42 & 348 & 536 & 52 & 96 & 0.33 & 0.60 & 2.51 & 0.83 \\  \hline
 3 & 78 & 1.0 & 13 & 0.16 & 0.51 & - & 493 & - & 78 & - & 0.49  & 1.77 & 0.86 \\  \hline
 4 & 79 & 5.6 & 21 & 0.51 & 1.41 & - & 75 & - & 38 & - & 0.24  & 0.23 & 0.65 \\  \hline
\end{tabular}%
\end{table}

\begin{flushleft} 
{\bf 3. Characterization of device 2}
\end{flushleft}

\begin{figure}[t]
\centering
\includegraphics[width=0.7\textwidth]{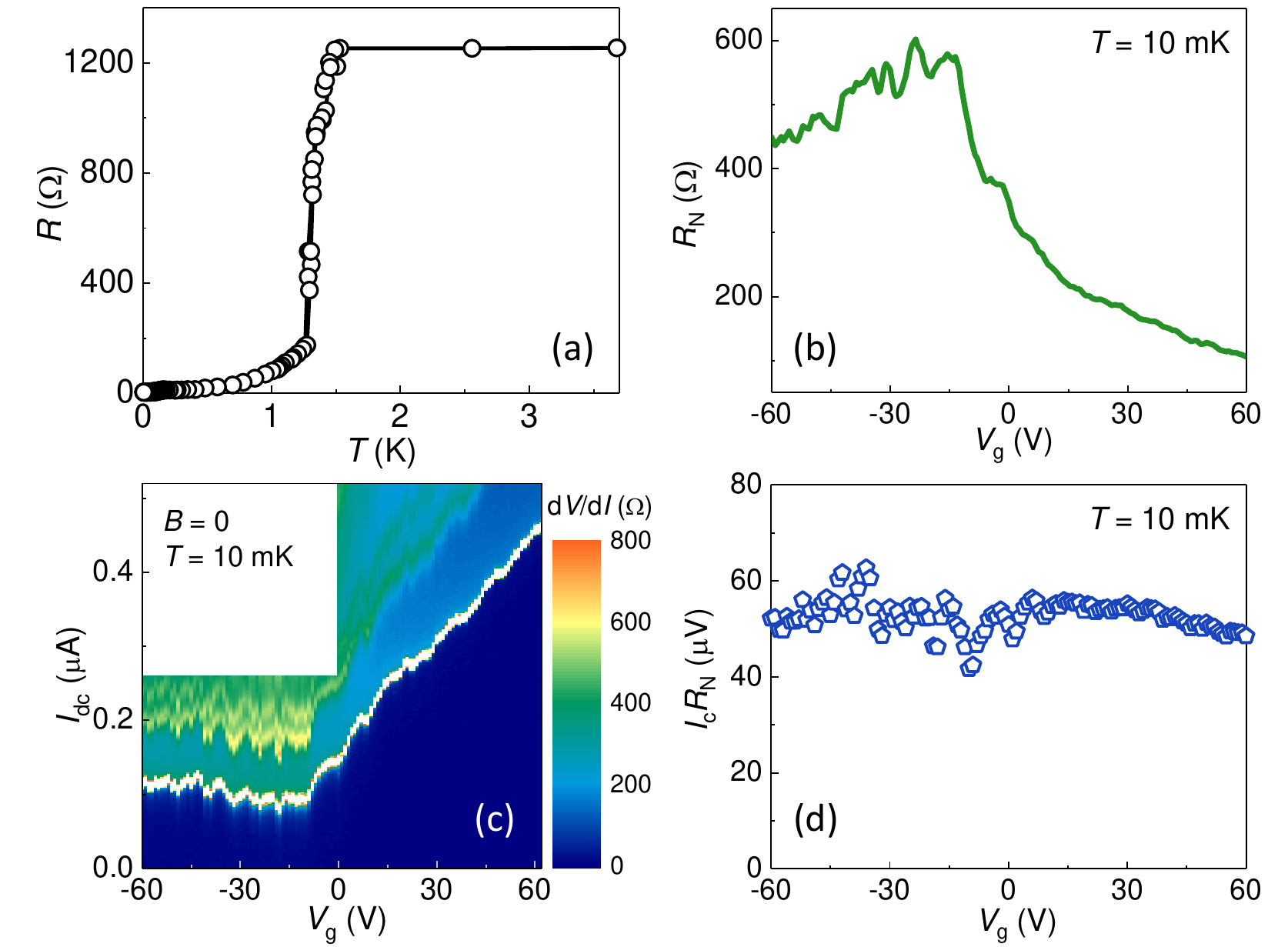}
\caption{(a) $R(T)$ curve of device 2 at $V_{\textrm{g}}=0$~V and $B=0$~mT. (b) Normal-state resistance  $R_{\text{N}}$ as a function of $V_{\textrm{g}}$ at $T=10$~mK. Due to the geometry of device 2, the resistance of the Al electrodes is added to the junction resistance when the Al electrodes are in the normal state; the $R$ value in (a) above 1.4 K includes this contribution. To avoid this contribution to affect our analysis, for the measurement of $R_\text{N}$ shown in (b) we employed a large ac excitation current $I_\text{ac}\gg I_\text{c}$ while keeping the Al electrodes in the superconducting state. The Dirac point is located at $V_{\textrm{g}} \approx -20$ V. The fluctuations in the $R_{\text{N}}(V_{\text{g}})$ curve was reproducible, which may arise from universal conductance fluctuations. (c) 2D plot of $\textrm{d}V/\textrm{d}I$ as a function of $I_{\text{dc}}$ and $V_{\textrm{g}}$ at $B=0$~mT. The critical current is represented by the white band at the edge of the superconducting region. (d) $I_{\text{c}}R_{\text{N}}$ product vs $V_{\text{g}}$ of device~2 at $T=10$~mK obtained from the data in panels (b) and (c).}
\end{figure}

Several Al-BiSbTeSe$_2$-Al Josephson junctions were measured down to low temperatures, and all investigated devices showed consistent behaviour. In Figure S2, we present a full characterization of device 2. Essentially all the results of device 1 are reproduced in device 2, except that the $E_{\textrm{F}}$ of device 2 is much closer to the Dirac point at $V_{\text{g}}=0$.

Similar to device 1, a two-step transition was seen in the cooling curve of device 2. The sharp transition at $T \approx 1.4\,$K  is associated with the superconducting $T_{\textrm{c}}$ of Al, and the broad transition occurring subsequently reflects the development of Josephson coupling. Due to the low carrier density at $V_{\text{g}}=0$, device 2 reaches zero resistance at a lower $T$ compared to device 1 (presented in the main text).
The Dirac point of device 2 is located at $V_{\textrm{g}} \approx -20\,$V (which we define as $V_{\rm DP}$), and the $E_{\text{F}}$ can be tuned deeply beneath the Dirac point by applying a large negative $V_{\textrm{g}}$, which enables us to investigate the $I_{\text{c}}$ vs $V_{\text{g}}$ relation in a wide range of the $p$-type region. The measured $I_{\text{c}}(V_{\text{g}})$ curve is illustrated by the white band in Figure S2(c). One can see that in the $p$-type region where $V_{\textrm{g}}<V_{\textrm{DP}}\approx-20$~V, the $I_{\text{c}}$ of device 2 actually increases slowly with decreasing $V_{\textrm{g}}$.
The $I_\text{c}R_{\text{N}}$ product of device 2 plotted in Figure S2(d) shows a behavior similar to that of device 1, although the minimum (at $V_{\textrm{g}} \approx -10$ V) is less clear.

\begin{flushleft} 
{\bf 4. Fraunhofer patterns of device 2 at various $V_{\textrm{g}}$}
\end{flushleft}

In Figure S3, we present the 2D colour maps of the $\text{d}V/\text{d}I$ data in the $I_\text{dc}$ vs $B$ plane measured in devices~2. Irregular Fraunhofer patterns with suppressed or missing lobes were observed in the entire range of the gate voltage. As $E_{\text{F}}$ approaches the Dirac point, the Fraunhofer pattern becomes blurred, and the oscillations of $I_{\text{c}}$ are only barely visible. This is presumably because the Josephson coupling is weakened at the Dirac point due to a low carrier density, and consequently the zero-resistance state is more prone to the thermal perturbation caused by electromagnetic noise.

\begin{figure}[h]
\centering
\includegraphics[width=0.95\textwidth]{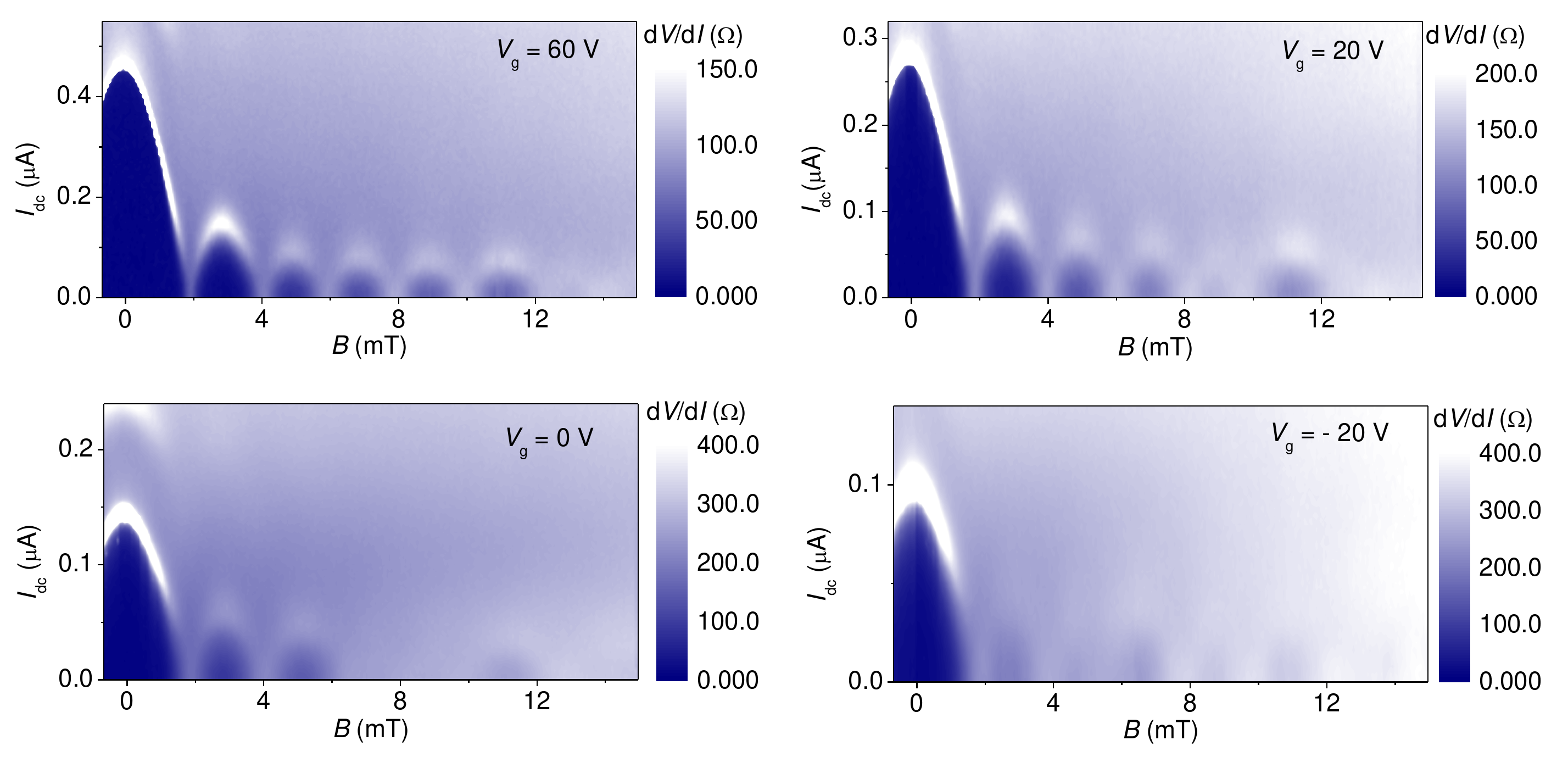}
\caption{2D plots of $\textrm{d}V/\textrm{d}I$ as a function of $B$ and $I_\text{dc}$, measured at $V_{\textrm{g}}$ = 60, 20, 0, and $-20$ V in device 2 at $T$ = 10 mK.}
\end{figure}

\begin{flushleft} 
{\bf 5. Estimation of Stewart-McCumber parameter}
\end{flushleft}

Figure S4 shows the $I$-$V$ curves of device~2 and device~3 measured in both forward and backward sweep directions. It is clear that the dc $I$-$V$ characteristics are non-hysteretic. In the RCSJ model, such behaviour is expected for overdamped Josephson junctions in which the Stewart-McCumber parameter $\beta_{c}$ ($\equiv 2\pi I_{\textrm{c}} (R_{\text{N}}^{I\textrm{-}V})^2 C / \Phi_0$) is much smaller than unity ($C$ is the shunt capacitance of the junction).
To calculate the $\beta_c$ of our junctions, we estimated the shunt capacitance $C$ by considering the two capacitors forming between the two Pt/Al/Au contact pads and the conductive Si substrate via the 290-nm-thick SiO$_2$ dielectric layer. It should be noted that these two capacitors are connected in series. Using the dielectric constant $\epsilon_{\text{r}}$ = 3.9 for SiO$_2$ and the areas of the contact pads of the devices, we  estimate $C \approx 10-30$ fF. The shunt capacitance originating from the direct capacitive coupling between the two closely spaced Al electrodes at the junction is estimated to be roughly three orders of magnitude smaller than this $C$ and hence is safely neglected.

Using the estimated values of $C$, we obtain $\beta_c \approx 0.2-2.5$ in different
devices (see Table S1). In the framework of the RCSJ-model, these junctions are at the boundary between the underdamped ($\beta_c \gg 1$) and overdamped ($\beta_c \ll 1$) regimes. In the literature, a previous study by Goldhaber-Gordon's group reported TI-based Josephson junctions in this regime, and a non-hysteretic behaviour similar to ours was observed \cite{Williams-PRL-2012}. The numerical calculation for junctions in this regime also gave similar non-hysteretic results \cite{JJ-book-Likharev}. Therefore, the absence of
hysteresis in the dc $I$-$V$ characteristics of our devices is consistent with their estimated $\beta_c$ values.

\begin{flushleft} 
{\bf 6. Estimation of the transparency of Al-BiSbTeSe$_2$ contacts}
\end{flushleft}

 The transparency $T_{\text{J}}$ of our Al-BiSbTeSe$_2$ superconducting contacts were estimated from the values of $I_{\text{e}}$ and $R_{\text{N}}^{I\textrm{-}V}$ by using the Octavio-Tinkham-Blonder-Klapwijk theory \cite{OTBK-1983,OTBK-1988}. Here, $I_{\text{e}}$ and $R_{\text{N}}^{I\textrm{-}V}$ are the excess current and the normal-state resistance extracted from the $I$-$V$ characteristics of the Josephson junctions, as defined in the main text. To identify $I_{\text{e}}$, the dc $I$-$V$ curves were measured to a high bias beyond $V_{\text{dc}}=2\Delta_{\text{Al}}/e$, where $2 \Delta_{\text{Al}} \approx 320\,\mu$eV is the superconducting gap of Al. The linear part of the high-bias $I$-$V$ curve was extrapolated to $V_{\text{dc}}=0$ and the intercept on the $I_{\text{dc}}$ axis gives $I_{\text{e}}$. The $R_{\text{N}}^{I\textrm{-}V}$ value was calculated as the inverse slope at high bias (i.e. $V_{\text{dc}}>2\Delta_{\text{Al}}/e$). The value of $\Delta_{\text{Al}}$ was obtained from the multiple Andreev reflection peaks as discussed in the main text. In Ref. \citenum{OTBK-1988}, the normalized excess current $I_{\textrm{e}}$$R_{\text{N}}^{I\textrm{-}V}$/$\Delta$ was numerically calculated as a function of the barrier strength $Z$, which is directly related to the transparency via $T_{\text{J}}=1/(1+Z^2)$. Using the given relation between $I_{\textrm{e}}$$R_{\text{N}}^{I\textrm{-}V}$/$\Delta$ and $Z$, we estimated $T_{\text{J}} \approx 0.65-0.86$ in our devices, as listed in Table S1.

\begin{figure}[h]
\centering
\includegraphics[width=0.7\textwidth]{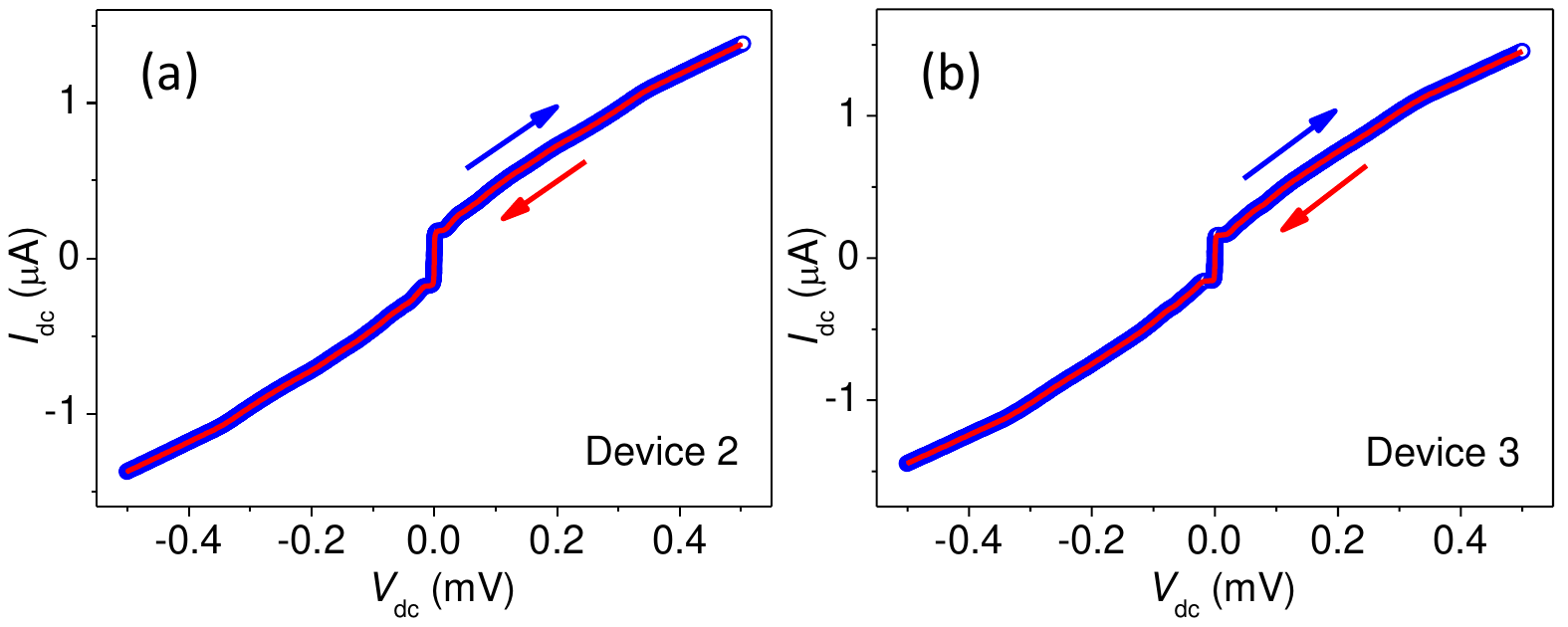}
\caption{$I$-$V$ characteristics of (a)~device~2  and (b)~device~3 measured in forward (blue circles) and backward (red lines) sweeps. Both curves fall on top of each other, indicating the absence of hysteresis in the dc $I$-$V$ characteristics. The wiggles in the curves are due to the multiple Andreev reflections discussed in the main text.}
\end{figure}

\begin{flushleft} 
{\bf 7. Estimation of $V_\text{CB}$}
\end{flushleft}

Considering the contributions from both top and bottom surfaces, the total two-dimensional carrier density $n_{\text{2d}}$ coming from the topological surface states in a ultra-thin sample is approximately given by
\begin{equation*}
n_{\text{2d}}(k)=2 \cdot \frac{1}{(2\pi)^2}\cdot \pi k^{2}=\frac{k^{2}}{2\pi}.
\end{equation*}
According to the ARPES results \cite{BSTS2-2012}, the conduction band bottom of BiSbTeSe$_2$ is located at $E_{\text{CB}}\approx0.2$~eV (measured from the Dirac point), and the Fermi wave vector of the surface states at this energy, $k_{\text{CB}}$, is about 0.076 \AA$^{-1}$. This gives $n_{\text{2d}}(k_{\text{CB}}) = 9.1 \times 10^{12}$ cm$^{-2}$.

The gate voltage required to tune $E_{\text{F}}$ from the Dirac point to the conduction band bottom is given by ($V_{\text{CB}}-V_{\text{DP}})=en_{\text{2d}}/C_{\text{g}}$. Here, $C_{\text{g}}=\epsilon_{0}\epsilon_{r}/d \approx 12$ nF/cm$^2$ is the capacitance per unit area of the back gate, with $\epsilon_{r}=3.9$ and $d=290$~nm being the dielectric constant and the thickness of the SiO$_2$ dielectric layer, respectively. With these values, we obtain $V_{\text{CB}}-V_{\text{DP}} \approx 121$ V. For $V_{\text{DP}}=-88$~V in device 1 shown in the main text, one obtains $V_{\text{CB}}\approx33$~V.

\begin{flushleft} 
{\bf 8. Applications of the maximum entropy fitting method}
\end{flushleft}

The presented anomalous Fraunhofer patterns have been analyzed using an advanced fitting scheme which is based on the maximum entropy principle. As a sanity check we applied the algorithm to an ideal pattern of a junction with a homogeneous current distribution given by $I_c(B)=I_\mathrm{c0} \left| \frac{\sin(\pi x)}{\pi x} \right|$, where $x=\Phi_J/\Phi_0$ and $\Phi_J$ is the flux through the junction. First, we initialized the current distribution to random numbers and then applied the same algorithm used to describe the anomalous patterns in the main text. Both the good fit to $I_c(B)$ as well as the flatness of the reconstructed current distribution (Figure S5(a)) show that the fluctuating current distribution extracted in the main text is neither due to over-fitting of the data, nor an artifact, but the most physical distribution explaining the experimental data.

We also applied the method to published data of SQUID oscillations measured from topological insulator Josephson junctions \cite{Kurter-2015} which have been argued by the authors to contain signatures of majorana bound states in the form of lifted SQUID nodes. To model the SQUID oscillations using our single junction fitting scheme, we forced the current distribution $j(0<x<W)$ to be zero in the interval $[w, W-w]$. Here, $w$ is the width of the individual junctions and $(W-2w)\cdot L_\mathrm{eff}=A_\mathrm{eff}$ represents the effective SQUID area with the effective length $L_\mathrm{eff}$ of the junctions. From the fit shown in Figure S5(b), we obtain $A_\mathrm{eff}=8.5\,\mu\mathrm{m}^2$, $W=13.3\,\mathrm{\mu m}$, $L_\mathrm{eff}=765\,\mathrm{nm}$ and $w=1.1\,\mathrm{\mu m}$. These values are consistent with the observed oscillation period and with the device geometry, taking into account the effect of magnetic flux focusing. The reconstructed current density $j(x)$ (inset of Figure S5(b)) shows moderate, yet finite spatial fluctuations whose amplitude and frequency are comparable to those discussed for our device in the main text. While the overall shape of the SQUID oscillations is well captured by our modeling, the lifting of the critical current close to the envelope's first node (marked by an arrow) cannot be fully accounted for. Even the tuning of the algorithm from maximum entropy towards least-square fitting does not improve the fit in this data range, which opens a room for discussing more involved mechanisms for the observed node lifting.

\begin{figure}[h]
\centering
\includegraphics[width=1\textwidth]{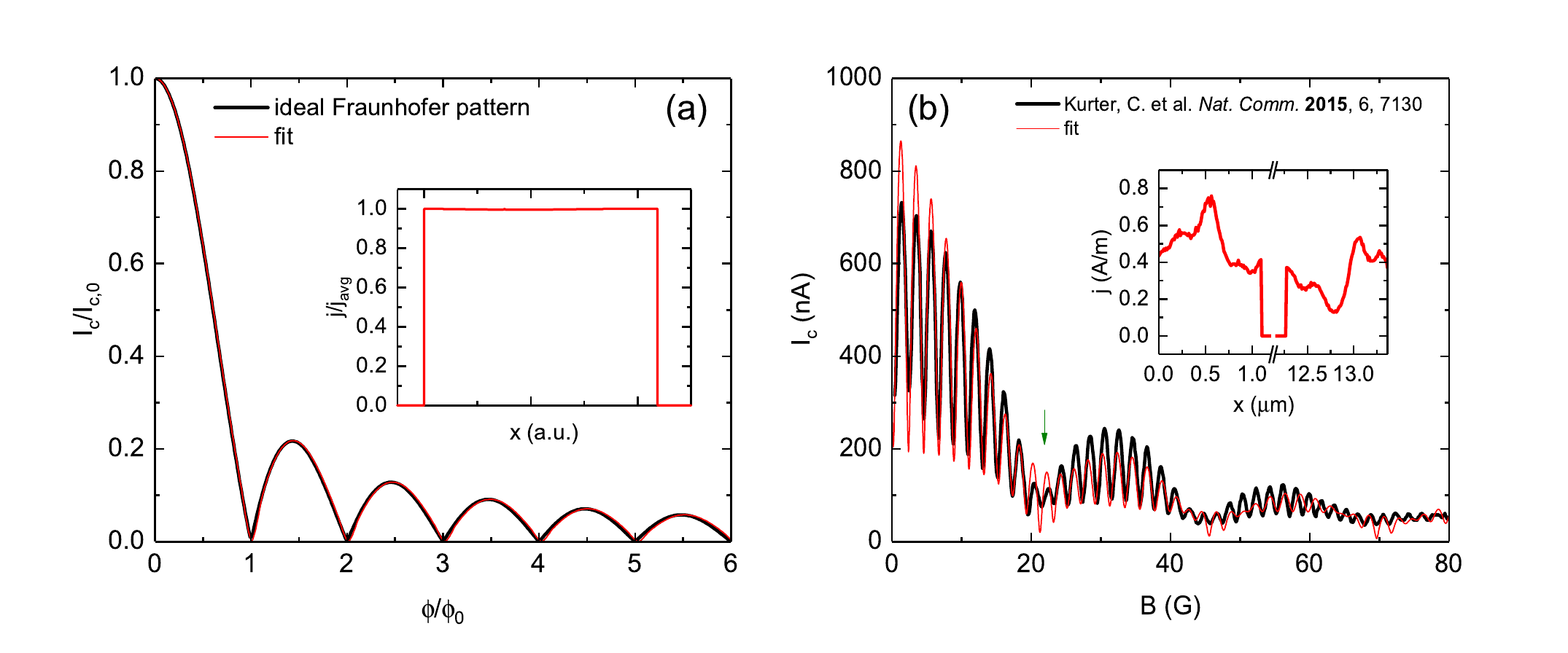}
\caption{(a) Fit (red line) of an ideal Fraunhofer pattern (black line) using the maximum entropy method. The obtained current density profile (inset) is completely flat and featureless. (b) Fit of the SQUID oscillations digitized from Ref.~\citenum{Kurter-2015}, Figure 2(a), ($V_\mathrm{TG}=-18$ V). The green arrow marks the oscillation envelope's first node where an unexpectedly finite critical current is observed. It is partly explained by a spatially fluctuating current density (inset). }
\end{figure}

\end{document}